\documentstyle[twoside,ijmpal,epsf]{article}

\newcommand{\beq}{\begin{equation}}
\newcommand{\eeq}{\end{equation}}
\newcommand{\bea}{\begin{eqnarray}}
\newcommand{\eea}{\end{eqnarray}}
\newcommand{\ol}[1]{\overline{#1}}

\newcommand{\mres}{m_{\rm res}}

\begin{document}


\normalsize\textlineskip
\thispagestyle{empty}
\setcounter{page}{1}

\vspace*{0.88truein}

\fpage{1}
\centerline{\bf LIGHT QUARK MASSES FROM QUENCHED LATTICE}
\vspace*{0.035truein}
\centerline{\bf QCD SIMULATIONS WITH DOMAIN WALL 
QUARKS\footnote{Talk given at DPF 2000, Columbus, OH, USA; work done in
collaboration with T.~Blum, P.~Chen, N.~Christ, C.~Cristian,
C.~Dawson, G.~Fleming, A.~Kaehler, X.~Liao, G.~Liu,
C.~Malureanu, R.~Mawhinney, S.~Ohta, G.~Siegert,
A.~Soni, C.~Sui, P.~Vranas, L.~Wu, and Y.~Zhestkov 
(RIKEN/BNL/CU Collaboration)
} }
\vspace*{0.37truein}
\centerline{\footnotesize MATTHEW WINGATE}
\vspace*{0.015truein}
\centerline{\footnotesize\it RIKEN BNL Research Center, 
Brookhaven National Laboratory}
\baselineskip=10pt
\centerline{\footnotesize\it Upton, NY 11973, USA}

\vspace*{0.21truein}
\abstracts{
Values for the strange quark mass and average up/down mass
have been obtained from quenched lattice QCD simulations
using the domain wall fermion action.  This discretization
preserves the properties of flavor and chiral symmetry at
nonzero lattice spacing.  Results are shown for 
two values of the lattice spacing.  The mass renormalization 
constant is computed nonperturbatively.
}{}{}


\textlineskip			
\vspace*{12pt}			

\noindent
The light quark masses are fundamental parameters of the
Standard Model and, as such, must be determined more precisely
in order to increase its predictive power.  Lattice QCD
can play an important role in computing the quark masses,
as it already has with the strong coupling constant.
Recently it was shown that the new domain wall fermion 
discretization can be applied to this endeavor.\cite{Blum:1999xi}
This talk reports on the progress of bringing these
calculations to the present level of other lattice
quark mass calculations.\cite{Gupta:DPF00,Lubicz:LAT00}

The RBC Collaboration has recently reported quenched
lattice QCD results for light hadron masses and
matrix elements from a series of simulations.\cite{Blum:2000kn}
We used domain wall quarks, which approximately maintain
continuum chiral symmetry by introducing
a fifth dimension of size $L_s$.\cite{Vranas:LAT00}
The nearly chiral surface
states at the boundaries mix within the bulk of the
fifth dimension.  This breaking effect can be parameterized 
for low energy matrix elements in terms of a residual quark 
mass $m_{\rm res}$.  A tunable quark mass $m_f$ is introduced
to the action which explicitly couples the surface states, thus
one studies operators containing quarks with effective masses
equal to $m_f + m_{\rm res}$.

\begin{center}
\begin{figure}[t]
\vspace{4.2cm}
\includegraphics{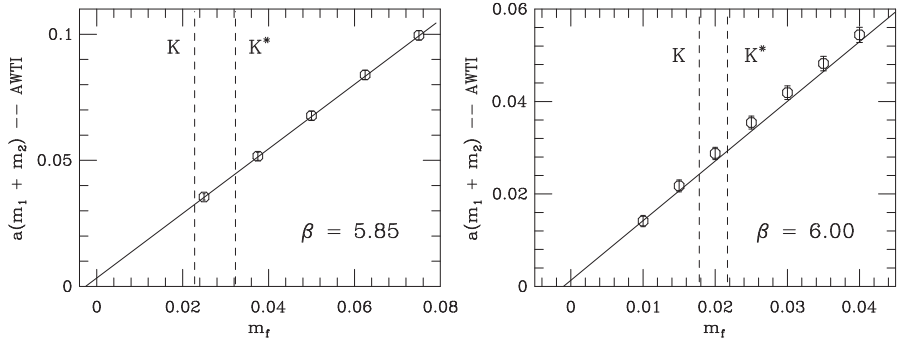}
\caption{
\label{fig:mq_pcac}
Axial WT identity quark mass (lattice units and regularization)
vs.\ $m_f$.  Dashed lines
indicate where the physical $K$ or $K^*$ mass is obtained.
}
\end{figure}
\end{center}
\begin{center}
\begin{figure}[t]
\vspace{4.2cm}
\includegraphics{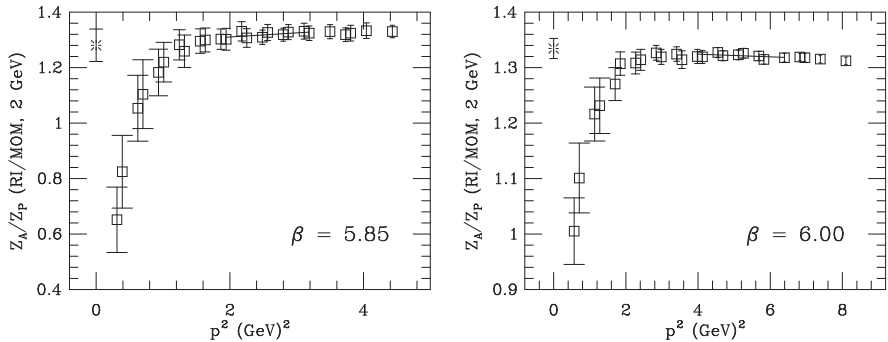}
\caption{
\label{fig:ZaZp_2gev}
$Z_A/Z_P$ in RI/MOM scheme at 2 GeV, plotted vs.\ input momentum
squared.  Line and asterisk indicate linear extrapolation
to $(ap)^2=0$ to remove any cutoff effects.
}
\end{figure}
\end{center}

\vspace{-1.5cm}
Here we report results from simulations using two
values of the gauge coupling parameter 
$\beta\equiv 6/g_0^2$.  
The lattice spacing is set through $M_\rho$.
The $m_f$ corresponding to the
physical strange sector, $m_f^{(s)}$, is set by requiring either the
pseudoscalar meson mass to equal $M_K$ or the vector mass to
equal $M_{K^*}$.  It is a common feature of quenched simulations
that $m_f^{(s)}$ from the $K$ is smaller than from $K^*$,
so results are quoted for both inputs.  The simulations here
are done with degenerate quarks, so 
$m_f^{(s)} \approx m_s/2$.  Simulations with nondegenerate quarks show
SU(3) flavor symmetry is preserved within errors.

In Fig.\ \ref{fig:mq_pcac} we show the quark mass
obtained through the axial Ward-Takahashi identity (AWTI)
by directly computing matrix elements of the local axial 
current and pseudoscalar density.  The renormalization
factor for this definition is $Z_A/Z_P$.  We compute
this nonperturbatively in the RI/MOM scheme\cite{Martinelli:1995ty} 
for several momenta; the results are plotted in 
Fig.\ \ref{fig:ZaZp_2gev}.\cite{Dawson:LAT00}

The quark mass may also be defined through the vector
Ward-Takahashi identity (VWTI).  The renormalization
constant for this definition is $Z_V/Z_S$.  Rather than
directly computing matrix elements, one can use the difference
$2(m_f^{(s)} - m_{\rm res}) = m_s + m_{u,d}$ (we 
assume\cite{Leutwyler:1996eq}
$m_{u,d} = m_s/24.4$).  Since the
vector current matrix element is an intermediate step, we
are free to choose the conserved one, for which $Z_V=1$.
The calculation of $Z_S$ is harder
since the wave function renormalization $Z_q$, which cancels
in the ratio $Z_A/Z_P$, now needs to be computed.  
So far, this has been done precisely only for $\beta=6.00$.

\begin{table}[t]
\tcaption{\label{tab:mstrange}
Strange quark mass in MeV ($\ol{\rm MS}, 2 ~{\rm GeV}$)
$\pm$ stat.\ $\pm$ sys.  Scale set through $M_\rho$.}
\centerline{\footnotesize
\begin{tabular}{lcccc}
{} 
 &\multicolumn{2}{c}{$K$ input}  
&\multicolumn{2}{c}{$K^*$ input}   \\ 
\multicolumn{1}{c}{$\beta$} & VWTI & AWTI & VWTI & AWTI \\ \hline
5.85   & -- & 100(5)(20) & -- & 138(7)(28) \\
6.00   & 110(2)(22) & 105(6)(21)  & 132(2)(26) & 127(6)(25) \\
\hline
\vspace{-0.7cm}
\end{tabular}
}
\end{table}

Table \ref{tab:mstrange} summarizes our results.  
The conversion to the $\ol{\rm MS}$
scheme was done perturbatively at two loops.  At 2 GeV
the three loop term is a sizable correction,\cite{Chetyrkin:2000pq}
which we include as a 5\% systematic error.  The only way
to reduce this uncertainty is to do the matching, and 
thus the whole simulation, with a scale closer to 3 GeV.
The $\approx 10\%$ systematic uncertainty from the lattice
spacing is not included, but should be if one uses
these values for phenomenology.

The results presented here are updated from 
last year.\cite{Wingate:2000yr}
The strange quark mass given then was taken using the 
VWTI, and a large uncertainty was assigned due to
the difference between $-\mres$ and the $m_f$ 
where $M_\pi^2\to 0$ linearly.  
Now that we have seen how
topological zero modes and quenching effects distort
the linear extrapolation to the chiral limit,\cite{Blum:2000kn}
this uncertainty is removed.


These are early steps in bringing
quenched domain wall calculations to the level of
those with other lattice actions.  Simulations with
finer lattice spacings are necessary to take the continuum limit,
and then an important check of systematic effects between
different discretizations can be made.

\nonumsection{Acknowledgments}

\noindent
Simulations done on the RBRC and CU QCDSP's
and the NERSC Cray T3E.

\nonumsection{References}

\end{document}